\documentclass[conference]{IEEEtran}
\IEEEoverridecommandlockouts
\usepackage{cite}
\usepackage{amsmath,amssymb,amsfonts}
\usepackage{algorithmic}
\usepackage{graphicx}
\usepackage{textcomp}
\usepackage{xcolor}
\usepackage{enumitem}
\usepackage{amsmath}
\def\BibTeX{{\rm B\kern-.05em{\sc i\kern-.025em b}\kern-.08em
    T\kern-.1667em\lower.7ex\hbox{E}\kern-.125emX}}
\begin{document}

\title{Security Enhancement of Quantum Communication in Space-Air-Ground Integrated Networks\\

\thanks{The work was supported by the  National Nature Science Foundation of China under Gants 62101450, in part by Shenzhen Science and Technology program under Grant GJHZ20220913143203006 and in part by Key R \& D Plan of Shaan Xi Province Grants 2023YBGY037.}
}
\author{
Author 1, Author 2, Author 3\\
Department of Something, University of Somewhere\\
City, Country\\
Email: \{author1, author2, author3\}@example.com
}

\author{
	\IEEEauthorblockN{
		Yixiao Zhang\textsuperscript{1,2}, 
		Wei Liang\textsuperscript{*1,2},
		Lixin Li\textsuperscript{1,2}, 
		Wensheng Lin\textsuperscript{1,2}
	} 
	\IEEEauthorblockA{\textsuperscript{1}School of Electronic and Information College, Northwestern Polytechnical University, China.}
	\IEEEauthorblockA{\textsuperscript{2}Research \& Development Institute of Northwestern Polytechnical University in Shenzhen, China\\ Email: zhangyixiao@mail.nwpu.edu.cn, \{liangwei, lilixin, linwest\}@nwpu.edu.cn}
}

\maketitle

\begin{abstract}
This paper investigates a transmission scheme for enhancing quantum communication security, aimed at improving the security of space-air-ground integrated networks (SAGIN). Quantum teleportation achieves the transmission of quantum states through quantum channels. In simple terms, an unknown quantum state at one location can be reconstructed on a particle at another location. By combining classical Turbo coding with quantum Shor error-correcting codes, we propose a practical solution that ensures secure information transmission even in the presence of errors in both classical and quantum channels. To provide absolute security under SAGIN, we add a quantum secure direct communication (QSDC) protocol to the current system. Specifically, by accounting for the practical scenario of eavesdropping in quantum channels, the QSDC protocol utilizes virtual entangled pairs to detect the presence of eavesdroppers. Consequently, the overall scheme guarantees both the reliability and absolute security of communication.
\end{abstract}

\begin{IEEEkeywords}
Space-Air-Ground Integrated Network, Quantum Teleportation, Quantum Secure Direct Communication, Turbo Coding
\end{IEEEkeywords}

\section{Introduction}
With the rapid development of technologies like the Internet of Things, cloud computing, and big data, SAGIN has become essential for global information connectivity. In the military field, the application of SAGIN technology is primarily reflected in integrated satellite-terrestrial communication systems, anti-jamming technologies, tactical data links, and advanced underwater communications\cite{RAY20226949}.
In civilian communication support, it involves satellite-based information relay applications, global mobile broadband, aviation management information, and maritime management information services. In civilian communication support, it involves satellite-based information relay applications, global mobile broadband, aviation management information, and maritime management information services. In commercial communication applications, SAGIN technology is widely used in mobile communication services, broadcasting, and in-flight internet access\cite{9312798}.
However, as the network system expands and its architecture becomes increasingly complex, the security threats faced by SAGIN are becoming more prominent. Issues such as unauthorized intrusions, illegal leakage of personal data, and inadequate defense of cybersecurity measures could potentially harm SAGIN communication systems. Therefore, addressing and resolving the security challenges of SAGIN is crucial\cite{9628162}.

Quantum communication technology, with its unique physical properties, has significant potential to enhance communication security. By leveraging quantum properties such as the no-cloning theorem and quantum entanglement, quantum communication can achieve unconditional secure communication, greatly improving overall security\cite{SHGC202408023}.
China has established several quantum communication backbone networks, including the "Beijing-Shanghai Line," "Wuhan-Hefei Line," and "Shanghai-Hangzhou Line." These networks use quantum key distribution technology to provide highly secure key distribution services for long-distance communication and ensure end-to-end secure communication in SAGIN through secure quantum keys\cite{chen2021integrated}.

In the post-quantum era, secure communication in SAGIN faces severe challenges. Currently, symmetric key encryption and public key encryption methods used in SAGIN are no longer secure with the advent of quantum computers\cite{DZYX202002001}. Quantum computers can solve large-scale NP-hard mixed-integer resource allocation problems in SAGIN\cite{9771632}, but integer factorization and discrete logarithm problems can be easily addressed by advanced quantum computers using computational power and quantum algorithms such as Shor's algorithm\cite{10437326}.

Nevertheless, quantum teleportation (QT) technology offers a secure protection scheme for data channels in SAGIN. QT encodes information into quantum state sequences and, through quantum entanglement, allows the receiver to accurately reconstruct the state of the original qubit without transmitting the qubit itself\cite{WLXB201723005}. QT has has been extensively evaluated for its potential applications in secure communication, quantum networks, and quantum repeaters\cite{hu2023progress}. The system is a dual-channel model consisting of a classical channel and a quantum channel. The classical channel transmits two classical bits. The quantum channel transmits one of the pre-shared entangled qubits from the sender to the receiver. The efficacy of QT is contingent upon a low-noise environment and the flawless transmission of both classical and quantum channels, whereas realistic error-correcting teleportation systems remain deficient.

This paper investigates a quantum communication security-enhanced transmission scheme for SAGIN. To address potential errors in both classical and quantum channels, a practical quantum teleportation scheme is proposed. The scheme ensures reliable transmission of measurement results by employing classical Turbo codes, while the transmission of pre-shared entangled qubits is protected using quantum error correction with Shor's code. Additionally, a trusted high altitude platform stations (HAPS) is used as the Einstein-Podolsky-Rosen (EPR) source, and entangled qubit pairs are verified through a quantum secure direct communication protocol. This explores a secure and reliable quantum teleportation protocol in the SAGIN environment.

The remainder of this paper is organized as follows: In Section II, we introduce the QT scheme under SAGIN, describing the classical and quantum channel models involved. In Section III, we present the improvement in Quantum Bit Error Rate (QBER) of the QT scheme with Turbo coding assistance, and analyze the impact of classical and quantum channel noise on QBER. Finally, we detail the steps for achieving an absolutely secure QT scheme using QSDC. Section IV concludes the paper.

\section{System Model and Problem Formulation}

Through the deep integration of satellite communication and HAPS with terrestrial wireless networks, non-terrestrial network communication overcomes terrain limitations and achieves ubiquitous coverage, especially in areas difficult to reach by traditional terrestrial networks, enabling seamless SAGIN coverage \cite{YDTX202307002}. 
Introducing quantum communication technology into SAGIN can significantly enhance overall network security. QT overcomes distance limitations caused by fiber and terrestrial channel losses. With the "Mozi" quantum science experimental satellite and space-based links, quantum state transmission has reached up to 1400 kilometers, laying the foundation for a global quantum internet \cite{ren2017ground}.

This paper proposes a communication scheme to implement QT in the SAGIN to enhance communication security. As shown in Fig. 1, the HAPS acts as a distribution center for entangled qubit pairs, providing a more stable and high-quality quantum communication channel between ground stations when communication is unstable due to terrain obstacles or other interferences. The HAPS generates entangled qubit pairs using an EPR source and distributes them to ground stations A and B. Upon receiving the entangled qubit pairs, ground station A performs a Bell-state measurement on its entangled qubit and the qubit to be transmitted. The measurement results (classical information) are then transmitted to ground station B via Low Earth Orbit (LEO) satellites. The LEO satellite serves as a relay for classical information, effectively reducing the propagation path of the signal through the atmosphere, thereby minimizing attenuation and improving communication quality. Upon receiving the classical information, ground station B performs the corresponding quantum operations on its entangled qubit based on this information, thereby reconstructing the original quantum state.
\begin{figure}[htbp]
\centerline{\includegraphics[width=0.45\textwidth]{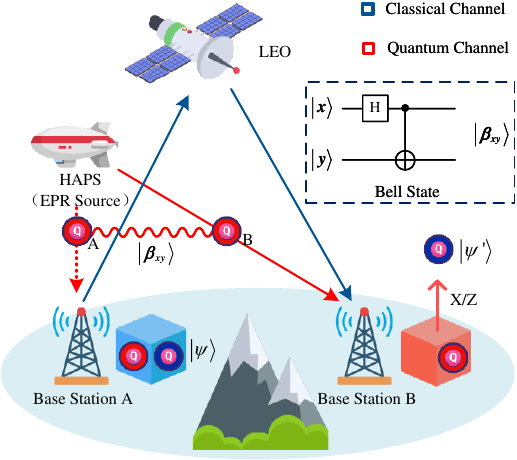}}
\caption{QT scheme under the SAGIN.}
\label{SAGIN}
\end{figure}

Where the EPR state is in a Bell state, which is used to describe the maximally entangled state of two qubits. In quantum circuit language, the method for generating a Bell pair between two qubits is shown in Fig. 1: first, apply a Hadamard gate to one of the qubits to convert it into the X-basis state, then apply a Controlled-NOT (CNOT) gate to the two qubits, with the qubit in the X-basis state serving as the control qubit.

For example, when the input is$\left| 00 \right\rangle $,applying the Hadamard gate transforms $\left| 0 \right\rangle $ into $H\left| 0 \right\rangle =(1/\sqrt{2})\left( \left| 0 \right\rangle +\left| 1 \right\rangle  \right)$, which is then used as the control input for the Controlled-NOT (CNOT) gate. The operational principle of the CNOT gate dictates that when the control qubit is in state $\left| 1 \right\rangle $, the target qubit experiences a NOT operation, resulting in a state flip; conversely, if the control qubit is in state 0, the target qubit remains unaffected. At this point, the input is in a superposition state $(1/\sqrt{2})\left( \left| 00 \right\rangle +\left| 10 \right\rangle  \right)$.The four Bell states are as follows:
\begin{equation}
\left| {{\beta }_{00}} \right\rangle =\frac{1}{\sqrt{2}}\left( \left| 00 \right\rangle +\left| 11 \right\rangle  \right)\label{eq}
\end{equation}
\begin{equation}
\left| {{\beta }_{10}} \right\rangle =\frac{1}{\sqrt{2}}\left( \left| 00 \right\rangle -\left| 11 \right\rangle  \right)\label{eq}
\end{equation}
\begin{equation}
\left| {{\beta }_{01}} \right\rangle =\frac{1}{\sqrt{2}}\left( \left| 01 \right\rangle +\left| 10 \right\rangle  \right)\label{eq}
\end{equation}
\begin{equation}
\left| {{\beta }_{11}} \right\rangle =\frac{1}{\sqrt{2}}\left( \left| 01 \right\rangle -\left| 10 \right\rangle  \right)\label{eq}
\end{equation}

\subsection{QT protocol}\label{AA}
The problem of QT can be described as follows: Alice and Bob are far apart. They share an EPR pair. When they separate, each takes one qubit of the EPR pair. Now, Alice wants to transmit a single qubit state $\left| \psi  \right\rangle $ to Bob. The state is $\left| \psi  \right\rangle =\alpha \left| 0 \right\rangle +\beta \left| 1 \right\rangle $.Alice does not kn.ow the state of this qubit and can only send classical information to Bob. The circuit for single-qubit teleportation is shown in the Fig. 2.

\begin{figure}[htbp]
\centerline{\includegraphics[width=0.4\textwidth]{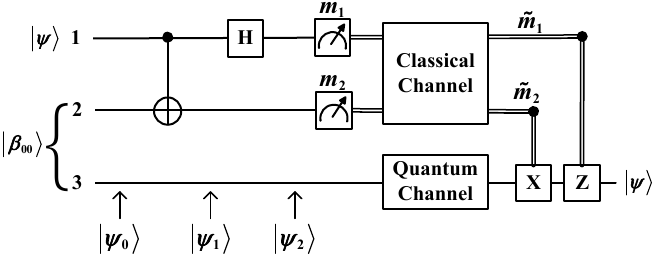}}
\caption{Single-Qubit teleportation circuit diagram.}
\label{fig2}
\end{figure}

At this point, the input state of the circuit is:
\begin{equation}
\begin{aligned}
   \left| {\psi_{0}} \right\rangle &= \left| \psi \right\rangle \otimes \left| {\beta_{00}} \right\rangle \\
   &= \left( \alpha \left| 0 \right\rangle + \beta \left| 1 \right\rangle \right) \otimes \frac{1}{\sqrt{2}} \left( \left| 00 \right\rangle + \left| 11 \right\rangle \right) \\
   &= \frac{1}{\sqrt{2}} \left( \alpha \left| 000 \right\rangle + \alpha \left| 011 \right\rangle + \beta \left| 100 \right\rangle + \beta \left| 111 \right\rangle \right)
\end{aligned}
\label{eq}
\end{equation}

In this setup, qubit 1 and qubit 2 belong to Alice, and qubit 3 belongs to Bob.

Next, Alice applies a CNOT gate to her qubit 2, with qubit $\left| \psi  \right\rangle $ as the control qubit. The circuit state at this point is:
\begin{equation}
\left| {{\psi }_{1}} \right\rangle =\frac{1}{\sqrt{2}}\left( \alpha \left| 000 \right\rangle +\alpha \left| 011 \right\rangle +\beta \left| 110 \right\rangle +\beta \left| 101 \right\rangle  \right)\label{eq}
\end{equation}

Then, Alice applies a Hadamard gate to qubit 1. The circuit state now becomes:
\begin{equation}
\begin{aligned}
  \left| \psi_{2} \right\rangle &= \left| 00 \right\rangle \frac{\alpha \left| 0 \right\rangle + \beta \left| 1 \right\rangle }{2} + \left| 01 \right\rangle \frac{\alpha \left| 1 \right\rangle + \beta \left| 0 \right\rangle }{2} \\
  &\quad + \left| 10 \right\rangle \frac{\alpha \left| 0 \right\rangle - \beta \left| 1 \right\rangle }{2} + \left| 11 \right\rangle \frac{\alpha \left| 1 \right\rangle - \beta \left| 0 \right\rangle }{2}
\end{aligned}
\label{eq}
\end{equation}
Here, $\alpha \left| 0 \right\rangle +\beta \left| 1 \right\rangle $is$\left| \psi  \right\rangle $, and Alice is in the state$\left| 00 \right\rangle $, meaning if Alice’s Bell measurement result is $\left| 00 \right\rangle $, then Bob’s qubit is in state $\left| \psi  \right\rangle $. For other Bell states, Bob needs to apply corresponding operations to recover the original unknown qubit state. See Table I for details.

\begin{table}[htbp]
\caption{Qubit 3 Gate Operations Based on Measurement Results}
\begin{center}
\begin{tabular}{|c|c|c|}
\hline
\textbf{Measurement Result} & \textbf{Qubit 3 State} & \textbf{Quantum Gate Operation} \\
\hline
0, 0 & $\alpha \left| 0 \right\rangle + \beta \left| 1 \right\rangle$ & / \\
\hline
0, 1 & $\alpha \left| 1 \right\rangle + \beta \left| 0 \right\rangle$ & X \\
\hline
1, 0 & $\alpha \left| 0 \right\rangle - \beta \left| 1 \right\rangle$ & Z \\
\hline
1, 1 & $\alpha \left| 1 \right\rangle - \beta \left| 0 \right\rangle$ & XZ \\
\hline
\end{tabular}
\label{tab1}
\end{center}
\end{table}

Next, this paper explores the effect of channel noise on teleportation. When only considering classical channel errors, the classical bits ${{\tilde{m}}_{1}}$ and ${{\tilde{m}}_{2}}$ can have four combinations: 00, 01, 10, and 11, as shown in Fig. 2. 
Only when both ${{m}_{1}}$ and ${{m}_{2}}$ are transmitted without errors can the receiver correctly enable or disable the X-gate and Z-gate. For example, if qubit 3 is in the state $\alpha \left| 0 \right\rangle +\beta \left| 1 \right\rangle $, the transmitted measurement result will be 00. In this case, an identity gate I should be applied at the receiver to obtain the final state $\left| \psi  \right\rangle $ (see Table I). However, if the received classical bit sequence is 01, an X-gate is incorrectly applied, resulting in the state:
\begin{equation}
\alpha \left| 0 \right\rangle +\beta \left| 1 \right\rangle \overset{X}{\mathop{\to }}\,\alpha \left| 1 \right\rangle +\beta \left| 0 \right\rangle \ne \left| \psi  \right\rangle\label{eq}
\end{equation}
This introduces a quantum flip error in the reconstructed qubit 1. 
Consequently, errors in the classical channel will result in quantum errors in the transmitted qubit 1. 

The Bit Error Rate (BER) is defined as the ratio of incorrectly received classical bits, denoted as ${{N}_{e}}$, to the total number of transmitted classical bits $N$:
\begin{equation}
\text{BER}=\frac{{{N}_{e}}}{N}\label{eq}
\end{equation}

Similarly, the QBER is given by:
\begin{equation}
\text{QBER}=\frac{N_{e}^{q}}{{{N}^{q}}}\label{eq}
\end{equation}
where ${{N}^{q}}$ represents the total number of transmitted qubits, and $N_{e}^{q}$ denotes the number of qubits received with errors.

For teleporting ${{N}^{q}}$ qubits, $N=2{{N}^{q}}$ classical bits must be transmitted to convey two measurement results. If a classical bit error \(N_{e}\) occurs, the worst-case scenario is a single classical bit error across two results, causing \(N_{e}^{q} = N_{e}\) quantum bit errors. The upper bound of the QBER is given by:

\begin{equation}
\text{QBER}=\frac{N_{e}^{q}}{{{N}^{q}}}=\frac{{{N}_{e}}}{0.5N}=2\text{BER}\label{eq}
\end{equation}

When considering quantum channel errors, the pre-shared qubit 3 might become a damaged qubit after transmission through a noisy quantum channel. To explore this scenario in more detail, suppose a qubit $\alpha \left| 0 \right\rangle +\beta \left| 1 \right\rangle $undergoes a quantum bit-flip (X) error during transmission. Then:
\begin{equation}
\alpha \left| 1 \right\rangle +\beta \left| 0 \right\rangle \overset{X}{\mathop{\to }}\,\alpha \left| 0 \right\rangle +\beta \left| 1 \right\rangle\label{eq}
\end{equation}
This will cause an error in the transmitted qubit 1.

The following part will analyze the probability of errors in the quantum channel ${{P}_{eq}}$. For example, if the error probability is ${{P}_{eq}}={{10}^{-1}}$, this means that on average, one in every ten pre-shared qubits is damaged. To quantify this phenomenon more precisely, let ${{N}^{q}}$ denote the total number of pre-shared qubits transmitted, and$\overset{-}{\mathop{N_{e}^{q}}}\,$ denote the number of damaged qubits. The quantum channel error rate can be expressed as:
\begin{equation}
P_{eq}^{~}=\frac{\overset{-}{\mathop{N_{e}^{q}}}\,}{{{N}^{q}}}\label{eq}
\end{equation}

QT is effective only if the noise level is low and both classical and quantum transmissions are error-free. Therefore, the subsequent steps will focus on improving the QBER of the teleportation protocol through classical channel coding and quantum channel coding.
\subsection{Classical Channel Model in SAGIN}

Due to the long transmission distance in satellite communication, the channel is often modeled using the Shadowed-Rician model. The communication link between base station A and base station B, via an LEO relay, can be represented by the Rician fading channel model :
\begin{equation}
H=\sqrt{\frac{{{p}_{0}}}{{{d}^{2}}}}(\sqrt{\frac{\zeta }{\zeta +1}}{{H}^{LOS}}+\sqrt{\frac{1}{\zeta +1}}{{H}^{NLOS}})\label{eq}
\end{equation}
where ${{p}_{0}}$ denotes the channel power loss at the reference distance, and $d$ is the communication link distance. $\zeta \ge 0$ represents the Rician factor, which specifies the power ratio between the Line-of-Sight (LOS) component and the Non-Line-of-Sight (NLOS) component of the channel. The LOS component of the channel is expressed as${{H}^{LOS}}$. ${{H}^{LOS}}$denotes the Rayleigh fading component of the channel.
\subsection{Quantum Channel Model in SAGIN}
In the proposed scheme for enhancing the security of SAGIN using QT, we employ the HAPS as an EPR source to distribute EPR pairs to base stations A and B. Given the proximity between the HAPS and base station A, we reasonably assume that the quantum channel loss during transmission to base station A is negligible, focusing only on the quantum channel loss between the HAPS and base station B. A depolarizing channel model is used for the simulation analysis.

The depolarizing quantum channel is characterized by three possible types of errors: bit-flip errors, phase-flip errors, and combinations of both (bit and phase flips occurring simultaneously). Specifically, bit-flip errors are equivalent to the application of a NOT gate (or Pauli-X gate), which is similar to classical bit flips. Phase-flip errors are equivalent to the application of a Z gate. Bit and phase-flip errors together are equivalent to the simultaneous application of X and Z gates. The depolarizing channel introduces X, Y, and Z errors with equal probability ${{p}_{e}}$. Therefore, the total error probability ${{P}_{eq}}$ is given by:
\begin{equation}
{{P}_{eq}}=3{{p}_{e}}\label{eq}
\end{equation}

In the quantum channel simulation, we refer to the channel modeling method described in \cite{8580740}. For each qubit, a random number $n$ (a floating-point number between 0 and 1) is generated, and operations are applied to the qubit according to the following rules.

\begin{itemize}
    \item If $\displaystyle n<\frac{1}{3}{{P}_{eq}}$, apply an X gate (bit-flip) to the qubit.\vspace{0.5em}
    \item If $\displaystyle \frac{1}{3}{{P}_{eq}}\le n<\frac{2}{3}{{P}_{eq}}$, apply a Z gate (phase-flip)  \vspace{0.5em} to the qubit. \vspace{0.5em}
    \item If $\displaystyle \frac{2}{3}{{P}_{eq}}\le n<{{P}_{eq}}$, apply a Y gate (both bit  \vspace{0.5em} and phase flip) to the qubit.\vspace{0.5em}
    \item If $\displaystyle n\ge {{P}_{eq}}$, no operation is applied to the qubit.
\end{itemize}

\section{Security-Enhanced QT Scheme Based on SAGIN}
\subsection{Turbo Coding-Assisted QT}\label{AA}
Our system selects a QPSK scheme to assist QT in the SAGIN. Fig. 3 shows the relationship between QBER/BER and signal-to-noise ratio (SNR) for the QPSK modulation scheme using Turbo coding. The effect on QBER is consistent with the previously explained relationship and is given by the following equation:
\begin{equation}
\text{QBER}\le 2\text{BER}\label{eq}
\end{equation}

Results indicate that introducing Turbo codes in the classical channel can improve the QBER of the QT protocol.

When the quantum channel is error-free, QBER is approximately twice the BER. However, with the introduction of an imperfect quantum channel, the upper bound of QBER can be expressed by the following formula:
\begin{equation}
\text{QBER}\le 2\text{BER}+P_{eq}^{~}\label{eq}
\end{equation}
This formula provides an upper limit since, in some cases, errors in the classical and quantum channels may offset each other, reducing the observed error rate.

Fig. 4 illustrates the QBER vs. SNR performance for a Turbo-coded 4-Phase Shift Keying (4-PSK) scheme in the SAGIN. As the SNR increases, the BER of the classical channel gradually decreases. It is observed that in the high SNR region, QBER reaches an error floor at $P_{eq}^{~}$, consistent with expectations.
\begin{figure}[htbp]
\centerline{\includegraphics[width=0.45\textwidth]{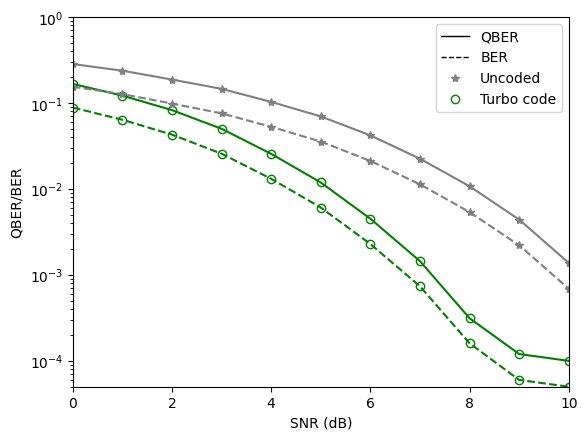}}
\caption{QBER/BER and SNR Performance Analysis: Comparison of communication performance between uncoded 4PSK and Turbo-coded 4PSK schemes in the SAGIN.}
\label{fig4}
\end{figure}

\begin{figure}[htbp]
\centerline{\includegraphics[width=0.45\textwidth]{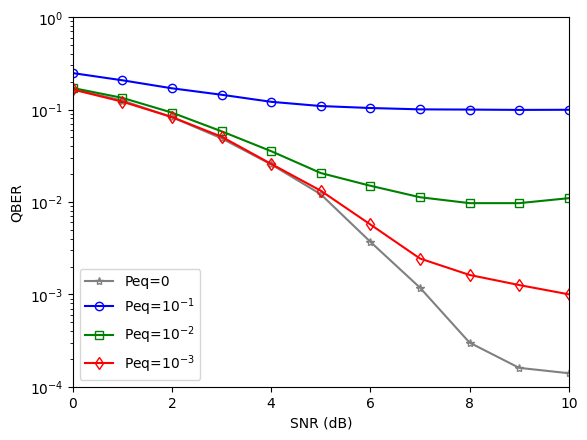}}
\caption{QBER and SNR Performance for Turbo-Coded 4PSK-Assisted Scheme in the SAGIN, with Quantum Channel Depolarization Probabilities$P_{eq}^{~}=({{10}^{-1}},{{10}^{-2}},{{10}^{-3}},0)$.}
\label{fig5}
\end{figure}
\subsection{Secure and Reliable Teleportation Protocol}
In the system model shown in Fig. 1, HAPS serves as the EPR source, with each qubit from the entangled pair being transmitted to Base Station A and Base Station B, respectively. This design provides additional security for the generation of entangled qubits and the transmission of EPR pairs. Once the entangled qubits are securely shared, QT can be considered absolutely secure. This is because the measurement results would only benefit an eavesdropper if they possessed qubit 3. The trusted HAPS enables QT to be applied as a one-time pad scheme in secure quantum communication \cite{Pirandola:20}.

However, even if EPR pairs are transmitted by trusted HAPS, there is a risk that qubits will be eavesdropped. The security of EPR pairs transmitted through quantum channels can be assessed based on the characteristics of quantum entanglement. Measuring a qubit in an entangled pair disrupts the entangled state, ultimately resulting in a corresponding pure state. Consequently, the interception of an EPR pair transmission by an eavesdropper will be detected instantaneously. If an eavesdropper intercepts a qubit and subsequently resends it after manipulation, the original structure of the entanglement is altered. Detection of such attacks can be achieved by measuring the transmitted and received qubits in the EPR pair and comparing the results \cite{PhysRevLett.84.4733}.

Consider the transmission of an EPR pair in which qubit A remains with the transmitter while qubit B is transmitted to the receiver. If an eavesdropper prepares the same EPR pair, say $\left| A{{B}^{00}} \right\rangle =(1/\sqrt{2})\left( \left| 00 \right\rangle +\left| 11 \right\rangle  \right)$, and captures qubit B, say  $\left| C{{D}^{00}} \right\rangle =(1/\sqrt{2})\left( \left| 00 \right\rangle +\left| 11 \right\rangle  \right)$, the system can be described as:
\begin{equation}
\begin{aligned}
  & \left| A{{B}^{00}} \right\rangle \left| C{{D}^{00}} \right\rangle =\frac{1}{2}(\left| A{{C}^{00}} \right\rangle \left| B{{D}^{00}} \right\rangle + \\ 
 & \left| A{{C}^{01}} \right\rangle \left| B{{D}^{01}} \right\rangle \left| A{{C}^{10}} \right\rangle \left| B{{D}^{10}} \right\rangle +\left| A{{C}^{11}} \right\rangle \left| B{{D}^{11}} \right\rangle ) \\ 
\end{aligned}
\label{eq}
\end{equation}
where,
\begin{equation}
\begin{aligned}
  & \left| i{{j}^{01}} \right\rangle =\frac{1}{\sqrt{2}}(\left| 01 \right\rangle +\left| 10 \right\rangle ) \\ 
 & \left| i{{j}^{10}} \right\rangle =\frac{1}{\sqrt{2}}(\left| 01 \right\rangle -\left| 10 \right\rangle ) \\ 
 & \left| i{{j}^{11}} \right\rangle =\frac{1}{\sqrt{2}}(\left| 00 \right\rangle -\left| 11 \right\rangle ) \\ 
\end{aligned}
\label{eq}
\end{equation}

Equation (19) indicates that if the eavesdropper measures the state \(\left| B{{D}^{00}} \right\rangle\), the remaining qubits will be in the state \(\left| A{{C}^{00}} \right\rangle\). Therefore, the original entangled state \(\left| A{{B}^{00}} \right\rangle\) is no longer valid, and qubits \(\left| A \right\rangle\) and \(\left| B \right\rangle\) are no longer entangled. When qubits are no longer entangled, one qubit's measurement result cannot be used to deduce the measurement result of another. If an eavesdropper is present, the QBER will be significantly elevated, which can be leveraged for secure quantum transmission.

This paper investigates a secure and reliable quantum transmission scheme based on QSDC using 9-qubit Shor encoding. The transmission process will be unconditionally secure, provided that the security of pre-shared entangled qubit pairs is ensured; thus, the protocol focuses on the security 
\begin{figure}[htbp]
\centerline{\includegraphics[width=0.45\textwidth]{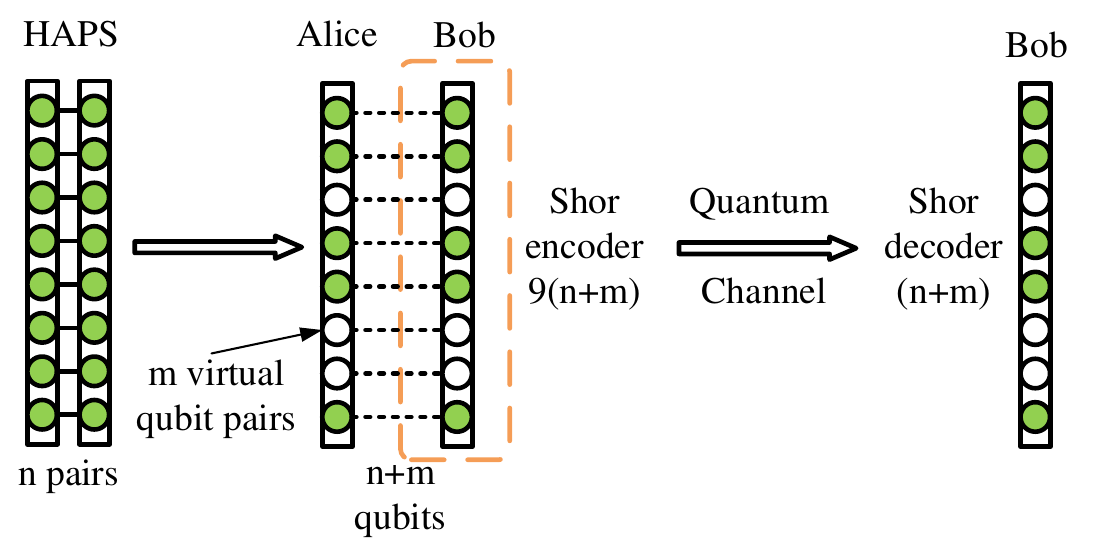}}
\caption{ Shor Encoding-Assisted QSDC.}
\label{fig6}
\end{figure}
of quantum channels. Furthermore, entangled pairs utilizing Shor decoding demonstrate greater reliability than uncoded schemes. The proposed secure and reliable transmission protocol, as illustrated in Fig. 5 , can be explained as follows:

\begin{enumerate}[label=(\arabic*)]
    \item The HAPS transmits one half of each of the $n$ EPR pairs to Alice and the other half to Bob. Each pair is in the state  $(1/\sqrt{2})(\left| 00 \right\rangle +\left| 11 \right\rangle )$.

    \item Next, $m$ virtual EPR pairs are inserted into secret positions within the original EPR qubit pairs. The virtual EPR pairs are in the state of $(1/\sqrt{2})(\left| 01 \right\rangle +\left| 10 \right\rangle )$, used for detecting eavesdroppers. As the value of $m$ increases, the protocol becomes more precise.

    \item There are now $(n + m)$ EPR pairs. The qubits to be sent to Bob are encoded using the Shor code at a 1/9 rate, generating $9(n + m)$ qubits.

    \item At Bob's side, the corresponding Shor decoding process is implemented to remove the redundant qubits and recover the $(n + m)$ qubits.

    \item Measurements on the decoded virtual qubits can be used to determine the extent of possible eavesdropping. The positions of the virtual qubits and the measurement results are transmitted to Alice. If there is no eavesdropper in the quantum channel, Alice's results should be opposite to Bob's, as the virtual qubits are in the state $(1/\sqrt{2})(\left| 01 \right\rangle +\left| 10 \right\rangle )$.

    \item If the QBER of the virtual qubits is below a chosen security threshold (which will be explained later), the pre-shared $n$ EPR pairs are considered secure, and the decoded EPR pairs can be used for transmission. However, if the QBER of the virtual qubits exceeds the threshold, indicating that the transmission has been intercepted, the entire process should be discarded, and the protocol should restart from step 1.

    \item When the EPR pairs are secure and reliable, information qubits (qubit 1) can be transmitted based on classical measurement bits.
\end{enumerate}

In step (6), a security error rate threshold must be set to compare with the error rate of the virtual qubits to determine if eavesdropping occurred during the transmission of qubits. This threshold needs to be established with the help of the Shor code. The security error rate can be specified from Fig. 6, where the x-axis corresponds to the 
\begin{figure}[htbp]
\centerline{\includegraphics[width=0.45\textwidth]{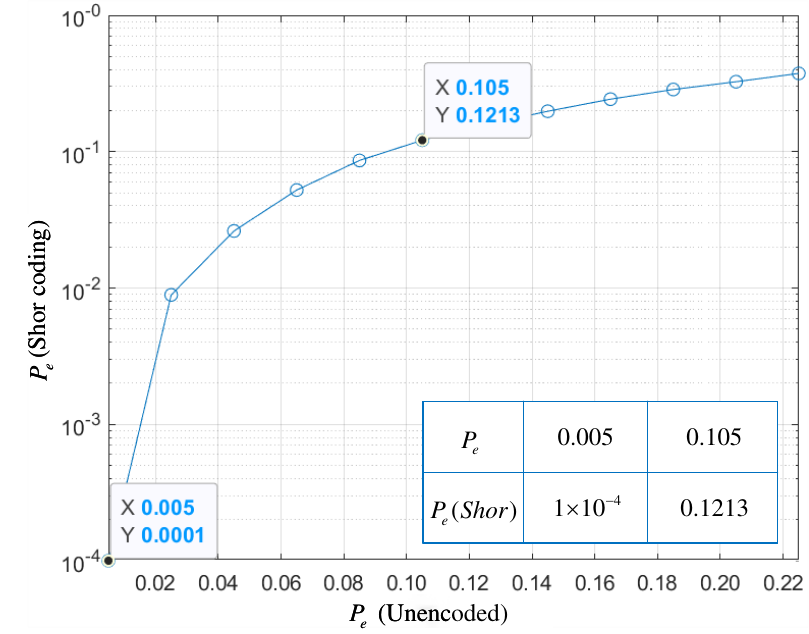}}
\caption{Comparison of Channel Depolarization Error Probability ${{P}_{e}}$  (Unencoded) and Error Probability ${{P}_{e}}(Shor)$ After Shor Decoding.}
\label{fig7}
\end{figure}
depolarization error probability in the quantum channel, ${{P}_{e}}$ and the y-axis represents the corresponding error rate after applying Shor decoding (denoted as ${{P}_{e}}(Shor)$).

Assuming no eavesdropper, the channel depolarization probability is ${{P}_{e}}=0.005$. It is reasonable to assume that an eavesdropper will augment the channel depolarization probability by a minimum of 10\%, resulting in an overall channel depolarization probability of ${{P}_{e}}>0.105$. From Fig. 6, after applying the Shor code, for instance, when ${{P}_{e}}=0.005$,  corresponds to ${{P}_{e}}(Shor)=1\times {{10}^{-4}}$, and when${{P}_{e}}>0.105$,  corresponds to${{P}_{e}}(Shor)>0.1213$. Thus, in cases when the quantum channel has a high depolarization error probability, it may be hard to identify the 10\% extra mistake caused by the eavesdropper in the absence of the Shor code. However, with the Shor code, the difference in QBER due to the presence of an eavesdropper becomes significantly larger, where ${{P}_{e}}(Shor)=1\times {{10}^{-4}}$ without an eavesdropper and ${{P}_{e}}(Shor)>0.1213$with an eavesdropper are distinct. In other words, the Shor code makes it easier to detect eavesdroppers. Additionally, when there is no eavesdropper, the reliability of the pre-shared quantum bits is significantly improved, increasing from ${{P}_{e}}=0.005$ to ${{P}_{e}}(Shor)=1\times {{10}^{-4}}$.
\section{Conclusion}
This paper investigates the performance of QT schemes assisted by classical Turbo codes and qubit Shor codes in the SAGIN. It investigates the secure transmission of pre-shared entangled qubits assisted by the Shor code based on the QSDC protocol. Specifically, when the quantum channel is imperfect, Shor codes are highly effective for detecting eavesdroppers. By applying Shor codes for the transmission of pre-shared quantum bits and Turbo codes for the classical transmission of measurement results, the proposed secure and reliable QT scheme can achieve ${{P}_{e}}(Shor)=1\times {{10}^{-4}}$ when the quantum channel depolarization probability reaches${{P}_{e}}=0.005$thereby enhancing communication security and reliability.

\vspace{12pt}

\end{document}